\documentclass[preprint,showpacs,preprintnumbers,amsmath,amssymb]{revtex4}

\usepackage{graphicx}
\usepackage{dcolumn}
\usepackage{bm}

\begin{document}

\preprint{APS/123-QED}

\title{The study of multifragmentation around transition energy in intermediate energy heavy-ion collisions\\}

\author{Karan Singh Vinayak}
\author{Suneel Kumar}%
 \email{suneel.kumar@thapar.edu}

\affiliation{%
School of Physics and Materials Science, Thapar University, Patiala-147004, Punjab (India)\\
}%
\date{\today}

\begin{abstract}

Fragmentation of light charged particles is studied for various systems at different incident energies between 50 and 1000 MeV/nucleon. We analyze fragment production at incident energies above, below and at transition energies using the isospin dependent quantum molecular dynamics(IQMD) model. The trends observed for the fragment production and rapidity distributions depend upon the incident energy, size of the fragments, composite mass of the reacting system as well as on the impact parameter of the reaction. The free nucleons and light charged particles show continous homogeneous changes irrespective of the transition energies indicating that there is no relation between the transition energy and production of the free as well as light charged particles.\\ 

\end{abstract}


\maketitle

\section{Introduction}
Nuclear Physics in general and heavy-ion collisions in particular are of central interest due to several rare phenomena emerging at different incident energies. Among these phenomena, collective flow and its disappearance\cite{1}, breaking of colliding nuclei into pieces, i.e. multifragmentation\cite{2}, subthreshold particle production\cite{3} as well as formation of the hot and dense nuclear matter\cite{4}  etc. have been discussed extensively in recent times. Another emerging area in the recent past is the isospin physics, which is now possible due to the availability of radioactive ion beams\cite{5}. One of its forms, directed transverse flow, is very sensitive toward physical scenario. At low incident energies, directed transverse flow is attractive that turns repulsive at higher incident energies. At a particular incident energy it disappears, known as balance energy. Another form of the flow, i.e. elliptical flow, also shows switch over from one region to other. The incident energy of this switch over is known as transition energy. The different positive or negative values of elliptical flow are related to the shape of the colliding matter. The elliptical flow describes the eccentricity of an ellipse like distribution. The positive value of elliptical flow reflects an in-plane emission, whereas out of plane is reflected by its negative value.\\

It is found that transition energy depends on the participant expansion and shadowing of the matter apart from the nucleons dominating the binary nucleonic collisions. Of course, all above listed phenomena are interrelated. Different flows represents collective velocity in a particular direction. Though, independent studies of various phenomena are available in the literature, no correlated study is  available in the literature that shows fragment structure at transition energy of elliptical flow. One is interested to know whether this structure differs from the multifragmentation obtained at below or above transition energy. Since transition energy is known for shape transition, it will be of interest to study where any non-trivial structure is obtained at transition energy or not. In earlier times, Puri and coworkers studied the fragmentation structure at balance energy and no particular pattern was observed at balance energy\cite{7}.\\ 
   
For the present study, isospin dependent quantum molecular dynamics(IQMD)\cite{5} model is used to generate the phase space of nucleons. Our present article is organized as follows. We discuss the model briefly in section II. Our results are presented in section III and we summarize the results in section IV.\\

\section{ISOSPIN-dependent QUANTUM MOLECULAR DYNAMICS (IQMD) MODEL}

The IQMD model\cite{5}, which is an improved version of the QMD model \cite{3,4,6} developed by J. Aichelin and coworkers, then has been used successfully to various phenomena such as collective flow, disappearance of flow, fragmentation \& elliptical flow. 
The isospin degree of freedom enters into the calculations via symmetry potential, cross-sections and
Coulomb interaction\cite{5}.
The details about the elastic and inelastic cross-sections
for proton-proton and neutron-neutron collisions can be found in Ref.\cite{5}. \\
In IQMD model, the nucleons of target and projectile
interact via two and three-body Skyrme forces, Yukawa potential and Coulomb interactions. 
In addition to the use of explicit charge states of all baryons and mesons, a symmetry potential between 
protons and neutrons corresponding to the Bethe- Weizsacker mass formula has been included.\\
The hadrons propagate using classical Hamilton equations of motion:
\begin{eqnarray}
\frac{d\vec{r_i}}{dt}~=~\frac{d\it{\langle~H~\rangle}}{d{p_i}}~~;~~\frac{d\vec{p_i}}{dt}~=~-\frac{d\it{\langle~H~\rangle}}{d{r_i}}
\end{eqnarray}
with
\begin{eqnarray}
\langle~H~\rangle&=&\langle~T~\rangle+\langle~V~\rangle\nonumber\\
&=&\sum_{i}\frac{p_i^2}{2m_i}+
\sum_i \sum_{j > i}\int f_{i}(\vec{r},\vec{p},t)V^{\it ij}({\vec{r}^\prime,\vec{r}})\nonumber\\
& &\times f_j(\vec{r}^\prime,\vec{p}^\prime,t)d\vec{r}d\vec{r}^\prime d\vec{p}d\vec{p}^\prime .
\end{eqnarray}
 The baryon-baryon potential $V^{ij}$, in the above relation, reads as:
\begin{eqnarray}
V^{ij}(\vec{r}^\prime -\vec{r})&=&V^{ij}_{Skyrme}+V^{ij}_{Yukawa}+V^{ij}_{Coul}+V^{ij}_{sym}\nonumber\\
&=&\left(t_{1}\delta(\vec{r}^\prime -\vec{r})+t_{2}\delta(\vec{r}^\prime -\vec{r})\rho^{\gamma-1}
\left(\frac{\vec{r}^\prime +\vec{r}}{2}\right)\right)\nonumber\\
& & +~t_{3}\frac{exp(|\vec{r}^\prime-\vec{r}|/\mu)}{(|\vec{r}^\prime-\vec{r}|/\mu)}
~+~\frac{Z_{i}Z_{j}e^{2}}{|\vec{r}^\prime -\vec{r}|}\nonumber\\
& &+t_{6}\frac{1}{\varrho_0}T_3^{i}T_3^{j}\delta(\vec{r_i}^\prime -\vec{r_j}).
\label{s1}
\end{eqnarray}
Here $Z_i$ and $Z_j$ denote the charges of $i^{th}$ and $j^{th}$ baryon, and $T_3^i$, $T_3^j$ are their respective $T_3$
components (i.e. 1/2 for protons and -1/2 for neutrons).
The parameters $\mu$ and $t_1,.....,t_6$ are adjusted to the real part of the nucleonic optical potential. 
For the density
dependence of nucleon optical potential, standard Skyrme-type parameterization is employed.
The potential part resulting from the convolution of the distribution function 
with the Skyrme interactions $V_{\it Skyrme}$ reads as :
\begin{equation}
{\it V}_{Skyrme}~=~\alpha\left(\frac{\rho_{int}}{\rho_{0}}\right)+\beta\left(\frac{\rho_{int}}{\rho_{0}}\right)^{\gamma}
~~\cdot
\end{equation}
The two of the three parameters of equation of state are determined by demanding that at normal nuclear matter density,
 the binding energy should be equal to 16 MeV. The third parameter $\gamma$ is usually treated as a free parameter.
 Its value is given in term of the compressibility:
\begin{equation}
\kappa~=~9\rho^{2}\frac{\partial^{2}}{\partial\rho^{2}}\left(\frac{E}{A}\right)~~\cdot
\end{equation}
The different values of compressibility give rise to soft(S) and hard(H) equations of state(EOS). It is worth mentioning that Skyrme 
forces are very successful in the analysis of low energy phenomena such as fusion, fission and cluster-radioactivity, where nuclear potential plays an important role in these phenomena\cite{8,chin}.\\
For the present analysis, a hard (H) equation of state
has been employed along with isospin dependent free nucleon-nucleon cross-section.\\
\section{Results and Discussion}
We here perform a complete systematic study of multifragmentation at, transition energy and shall compare outcome at incident energies below and above the transition energies. Our results are  spanned over entire periodic table, and in particular, we simulated the results of $_{20}Ca^{40}~+~_{20}Ca^{40}$, $_{54}Xe^{131}~+~_{54}Xe^{131}$ and $_{79}Au^{197}~+~_{79}Au^{197}$.\\

\begin{figure}
\includegraphics[scale=0.42]{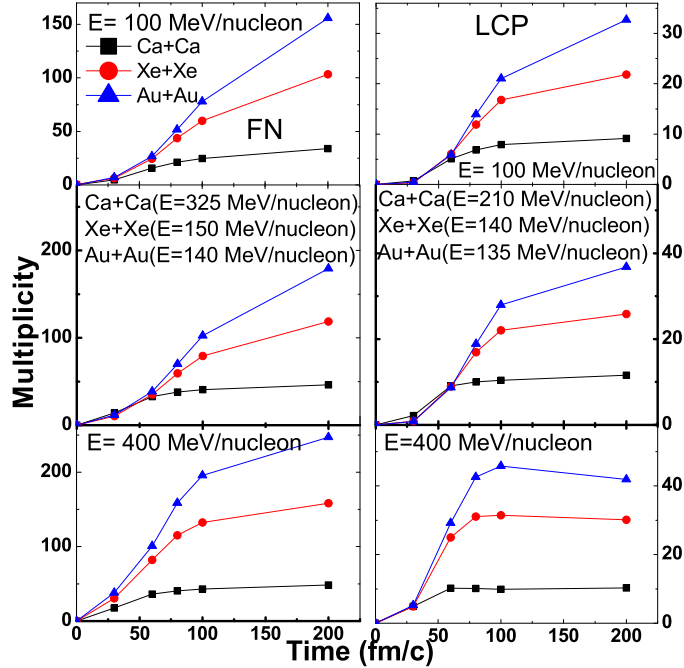}
\caption{\label{fig:1} The time evolution of multiplicity for free nucleons(F.N.'s)[A = 1](left)and
light charged particles (LCP's)[2$\le$ A $\le$ 4](right). The top, middle and bottom panel represents the incident beam energy i.e. at below transition,at transition and above transition energy respectively.}
\end{figure}

 The phase space generated by IQMD model  has been analyzed using minimum spanning tree(MST) method\cite{8} and analysis packages\cite{6}. Two nucleons are bound in a fragment if their distance is less than 4 fm. We also understand that in recent times, more sophisticated algorithm are also available in the literature\cite{6}. Since our phase space is analyzed at 200 fm/c, we assume that MST method should be able to detect the true fragmentation structure. As stated above, elliptical flow vanishes at a particular energy known as transition energy. In our study, elliptical flow is defined as the difference between the major and minor axes as :
Mathematically, it can be written as  

\begin{equation}
v_2~=~ <cos2\phi>~=~\langle\frac{p_x^2 - p_y^2}{p_x^2 + p_y^2}\rangle,
\end{equation}\\

where $<$$p_{x}$$>$ and $<$$p_{y}$$>$ are the x and y components of  momentum. The 
$<$$p_x$$>$ is in the reaction plane, while, $<$$p_y$$>$ is 
perpendicular to the reaction plane and $\phi$ is the azimuthal angle of the emitted particles momentum relative to  x-axis.
A positive value of the elliptical flow shows a rotational behaviour of the matter. The zero value represents the isotropic distribution in the transverse plane. As reported in the ref.\cite{5}, the transition energies for the reaction $_{20}Ca^{40}~+~_{20}Ca^{40}$, $_{54}Xe^{131}~+~_{54}Xe^{131}$ and $_{79}Au^{197}~+~_{79}Au^{197}$, are 325, 150, 140 MeV/nucleon, respectively, for free nucleons and 210, 140, 135 MeV/nucleon for light charged particles respectively.
 We shall, therefore, simulate the above reactions at these transition energies as well as at incident energies below and above transition energies.

\begin{figure}
\includegraphics[scale=0.40]{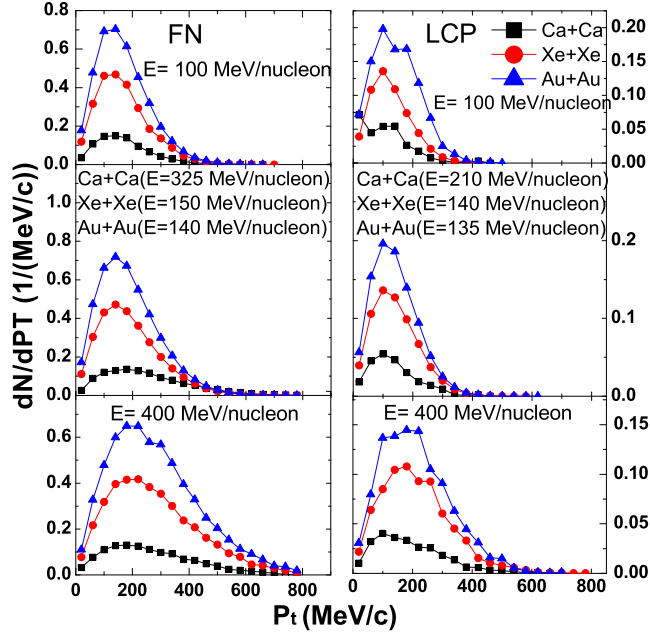}

\caption{\label{fig:2} Variation of number of particles with transverse momentum for F.N.'s(left) and L.C.P.'s(right). The different panels have same meaning as that of Fig. 1.}
\end{figure}
   
In Fig.1, time evolution of free nucleons (FN's)[A = 1] as well as light charged particles (LCP's)[2 $\leq$ A $\leq$ 4] is displayed corresponding to the reactions of $_{20}Ca^{40}~+~_{20}Ca^{40}$, $_{54}Xe^{131}~+~_{54}Xe^{131}$ and $_{79}Au^{197}~+~_{79}Au^{197}$. We also display the results at incident energies of 100 MeV/nucleon (below transition energy) as well as at 400 MeV/nucleon(for above the transition energy). A steady increase is observed in the production of free particles as well as LCP's with incident energies. We did not see of particular behaviour at corresponding transition energies. We see that the reaction is over quite early in lighter colliding nuclei compared to heavier called nuclei where nucleonic interactions keep going for larger times. In these cases, reactions are not over even at 200 fm/c. No typical behaviour is observed  at transition energies.
In addition to the mean field and NN collisions, other effects like the expansion of matter above the transition energy and shadowing  of the spectator below the transition energy take place. One should note that free nucleons result from the  binary collision whereas for the  production of LCP's, mean field play dominant role.

In fig.2, we  display the number of particles in a particular transverse momentum region. The transverse momentum is a factor that is used in the denominator of the elliptical flow. The transverse momentum is equal to the sum of the squares of momentum in x and y directions assuming the reaction plane to be the xz plane. Again, it is clear from the figure that with the increase in the incident energy from below to above the transition energy, nucleon with larger transverse energy starts dominating the  physics. Nucleons with high $p_{t}$  tail start emerging indicating that highly energetic nucleon appears at incident energies above transition energy.\\
 
To observe nucleonic stopping,  we plot  rapidity distribution in fig.3. It is observed that fragments are insensitive towards the nuclear stopping. This is in agreement with ref.\cite{chin}. Moreover, nuclear stopping is found to vary above and below the transition energy. The rapidity distribution denotes the nuclear stopping in term of Gaussians.\\ 
\begin{figure}
\includegraphics[scale=0.40]{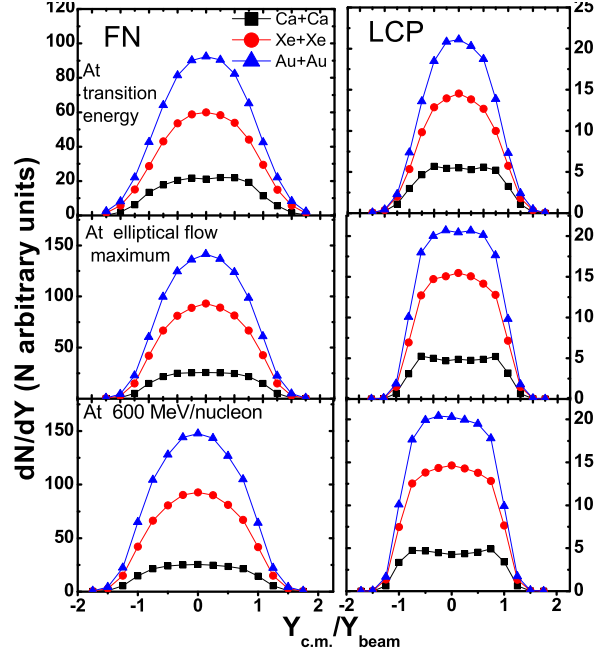}
\caption{\label{fig:3} Rapidity distribution of Free Nucleons[A = 1](left) and Light charged particles[2 $\le$ A $\le$ 4](right). The middle panel represent the rapidity distribution at energy at which there is maximum elliptical flow. The top and bottom panel represent rapidity distribution at transition energy and at 600 MeV/nucleon, respectively.}
\end{figure}

Narrower is the Gaussian, more is nuclear stopping. From the graph, it is also clear that Gaussian is narrower for the energy at which elliptical flow is maximum and starts becoming broader below and above this particular energy of maximum elliptical flow.\\

One can further say, that for nuclear stopping, the energy at which elliptical flow is maximum is more important  compared to the values at which transition takes place from positive to negative values i.e., changing from rotational to expansion behaviour. Interestingly, Gaussian's are more narrower for free particles  compared to LCP's.  It is also shown by us \cite{chin} that free particles  fail to explain some of the properties of nuclear stopping, while all properties can be explained by the LCP's in a reasonable manner.\\  

\begin{figure}
\includegraphics[scale=0.40]{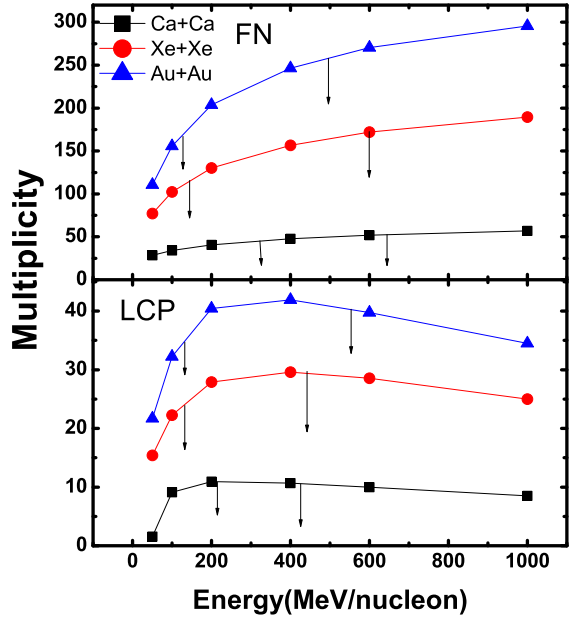}
\caption{\label{fig:4} Incident energy dependence of multiplicity for different symmetric systems. The top and bottom panel represents the F.N.'s and L.C.P.'s, respectively.}
\end{figure}

From the above discussion for figs.1 and 2, it becomes quite important to see the incident energy dependence of the multifragmentation( i.e. for F.N. and LCP's). Once again, different behaviour of fragments is observed with  different incident energies (as shown in fig. 4) . The free particles are found to increase with the incident energy while LCP's are found to decrease after certain incident energy. The change in the behaviour is also true for elliptical flow and nuclear stopping.\\ 

It  indicates that the study of LCP's, below and above as well as at  transition energy is more important  compared to free particles. The arrows are indicating the values for different systems at which transition takes place as well as at values corresponding to the maximum elliptical flow. From the lower panel, it is observed that as soon as the elliptical flow become maximum, the production of the LCP's start varying drastically. From here, we conclude the resemblence in the behaviour of LCP's, nuclear stopping and elliptical flow, which can be the backbone of heavy-ion physics.\\  

\begin{figure}
\includegraphics[scale=0.40]{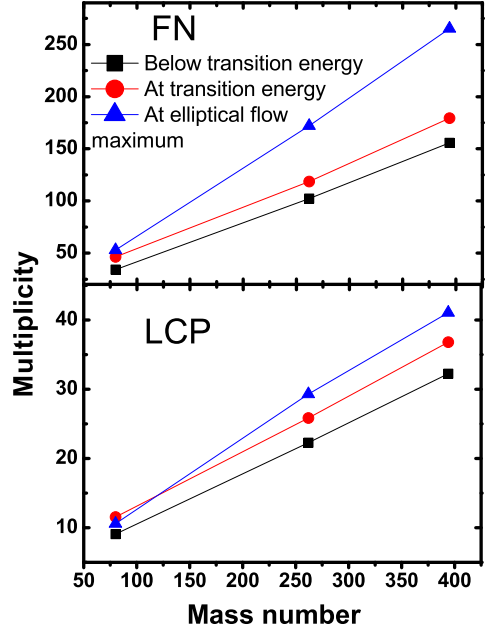}
\caption{\label{fig:5} Multiplicity as a function of the composite mass of the system. The upper panel is for the free nucleons, while lower panel is for the L.C.P.'s.}
\end{figure}

At last, we have tried to fit the multiplicity of free particles and light charged particles (LCP's) with the total mass of the system at corresponding transition energies as well as at incident energies below and above the transition energies. The free and LCP's are found to increase with the increase in the mass number of the system. The main point to be noted is that the variation in the free particles with the variation in the energy is more  compared to light charged particles. 

This is also one of the important points in the present study. The reason is that the free particles originates from the participant zone, which keeps increasing with the increase in the incident energy.\\

On the other hand,  LCP's have also some contributions from the spectator matter in contrast to the free nucleons that emerges from the participant region only. Due to this reason, the difference in the production of LCP's is less  compared to the production of the free particles. We also fit the multiplicity in terms of the mass number. The multiplicity is parametrized as the power law of the form $A^{\tau}$. The values of $\tau$ in case of free nucleons at different incident energies i.e. below, at and above transition energy are 0.97, 0.88 and 1.02 respectively. Similarly, the values of $\tau$ in case of LCP's are 0.81, 0.74 and 0.84 at the three different incident energies i.e. below, above and at transition energy respectively. The variation in the values of $\tau$ is more in case of free nucleons compared to light charged particles.


\section{Conclusion}
In Conclusion, we have investigated  various properties of multifragmentation for different reacting systems at incident energies between 50 and 1000 MeV/nucleon using the isospin dependent quantum molecular dynamics(IQMD) model. This study was conducted with an aim to see whether there is any particular structure in the fragmentation at and around transition energies or not. Our detailed study  over wide range incident energies shows no typical structure at transition energies. We observe a continous homogeneous change in the production of light charged particles and free nucleons.



\end{document}